\documentclass[letterpaper]{article} 
\usepackage{aaai2026}  
\usepackage{times}  
\usepackage{helvet}  
\usepackage{courier}  
\usepackage[hyphens]{url}  
\usepackage{graphicx} 
\usepackage{natbib}  
\usepackage{amssymb}
\usepackage{amsmath}
\usepackage{booktabs}
\usepackage{multirow}
\usepackage{caption} 
\usepackage{algorithm}
\usepackage{algorithmic}

\usepackage{newfloat}
\usepackage{listings}
\DeclareCaptionStyle{ruled}{labelfont=normalfont,labelsep=colon,strut=off} 
\floatstyle{ruled}
\newfloat{listing}{tb}{lst}{}
\floatname{listing}{Listing}
\title{HyMoERec: Hybrid Mixture-of-Experts for Sequential Recommendation \\ (Student Abstract)}
\author{
    Kunrong Li\textsuperscript{\rm 1}, Zhu Sun\textsuperscript{\rm 1}, Kwan Hui Lim\textsuperscript{\rm 1}
}
\affiliations{
    \textsuperscript{\rm 1}Singapore University of Technology and Design, Singapore\\

    kunrong\_li@mymail.sutd.edu.sg, zhu\_sun@sutd.edu.sg, kwanhui\_lim@sutd.edu.sg
}

\usepackage{bibentry}

\begin{document}

\vspace{-10pt}
\maketitle

\vspace{-10pt}
\begin{abstract}
We propose \textbf{HyMoERec}, a novel sequential recommendation framework that addresses the limitations of uniform Position-wise Feed-Forward Networks in existing models. Current approaches treat all user interactions and items equally, overlooking the heterogeneity in user behavior patterns and diversity in item complexity. HyMoERec initially introduces a hybrid mixture-of-experts architecture that combines shared and specialized expert branches with an adaptive expert fusion mechanism for the sequential recommendation task. This design captures diverse reasoning for varied users and items while ensuring stable training. Experiments on MovieLens-1M and Beauty datasets demonstrate that HyMoERec consistently outperforms state-of-the-art baselines.
\end{abstract}

\section{Introduction}

Sequential recommendation is a core task in modern recommendation systems, which predicts the next item based on historical behavior. 
Recent advances, ranging from recurrent neural networks~\cite{jannach2017recurrentgru4rec}, attention-based architectures~\cite{sun2019bert4rec,li2017neuralnarm}, to state-space models~\cite{liu2024mamba4rec} and large language models~\cite{li2025rallm}, have greatly improved sequence modeling in this domain.
However, a critical yet underexplored aspect lies with the Position-wise Feed-Forward Network (PFFN), which is the primary non-linear transformation module in most models. It applies the same FFN operation for all tokens and treats each item equally. 
This design neglects two key challenges: (1) User behavior heterogeneity: Users follow varied sequential patterns, with some interactions captured by simple dependencies and others requiring complex reasoning, which a uniform FFN cannot adaptively model; (2) Item complexity diversity: While popular items may be predicted from broad patterns, niche items often demand specialized representations, making identical computational pathways insufficient and less effective.

\begin{figure}[t]
    \centering
    \includegraphics[width=0.85\linewidth]{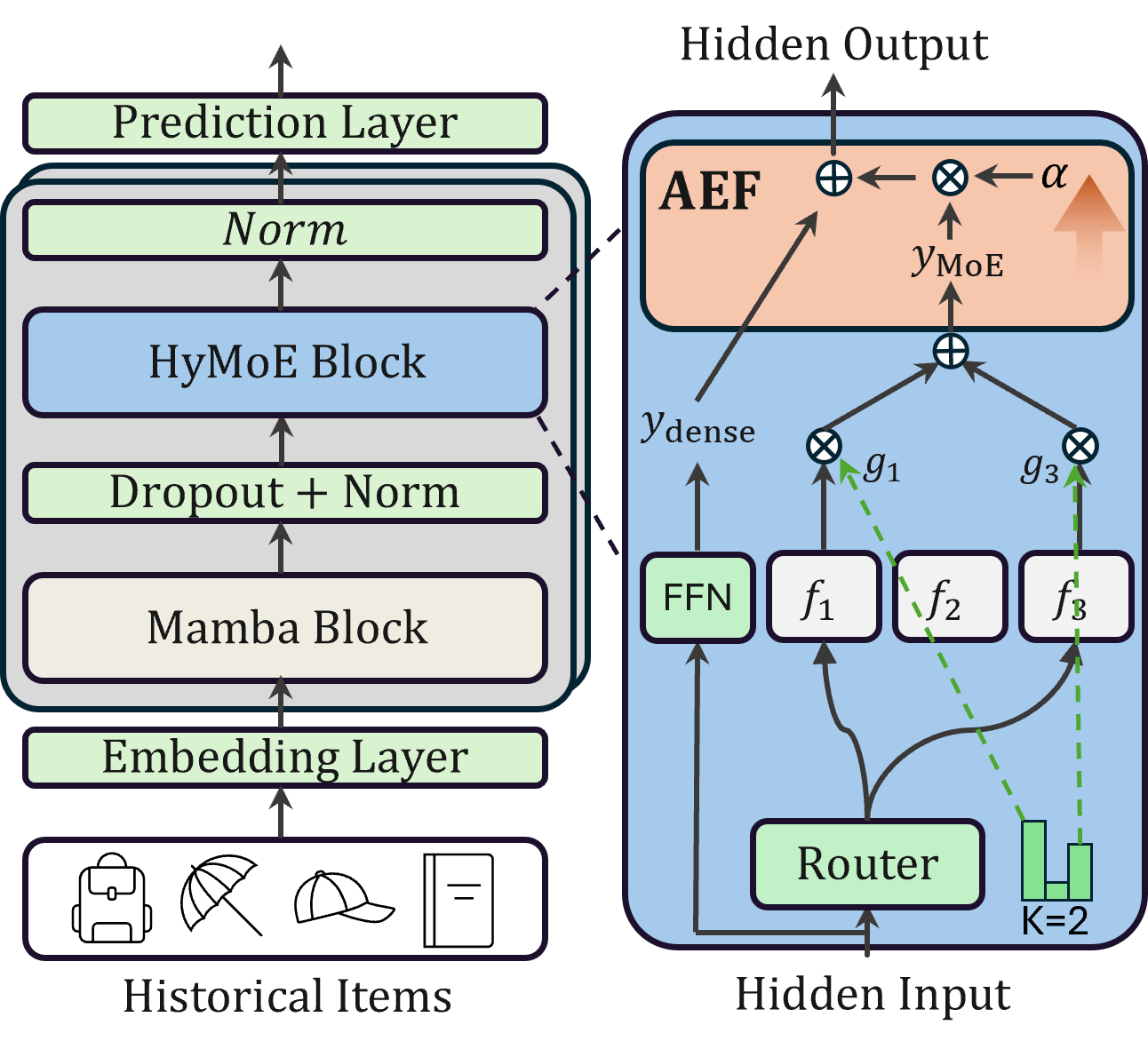}
    \caption{The overall framework of HyMoERec.}
    \label{fig:placeholder}
    \vspace{-10pt}
\end{figure}

To address these challenges, we propose \textbf{HyMoERec}, a sequential recommendation framework that redefines PFFN processing via a hybrid mixture-of-experts architecture, which is illustrated in Fig. \ref{fig:placeholder}. Our approach introduces three innovations: (1) \textit{Hybrid Mixture of Experts (HyMoE)}: A combination of one shared expert and multiple specialized experts. Unlike conventional MoE relying solely on sparse selection, this hybrid design ensures stable optimization through consistent learning signals while enabling specialization for domain-specific patterns; (2) \textit{Adaptive Expert Fusion (AEF)}: A learnable gating parameter $\alpha$ that dynamically balances shared and specialized pathways. A progressive warm-up schedule begins with shared expert reliance and gradually integrates specialized knowledge, while load-balance regularization promotes diverse expert usage; (3) \textit{Empirical validation}: Comprehensive experiments in MovieLens-1M and Amazon Beauty demonstrate that HyMoERec consistently exceeds state-of-the-art baselines.

\section{Methodology}

\textbf{Hybrid Mixture-of-Experts (HyMoE).} A common limitation in sequential recommendation is that all tokens are processed by a uniform PFFN, regardless of their semantic or temporal characteristics. This uniformity underutilizes the model capacity and fails to capture the heterogeneity of user behaviors.
Therefore, we introduce Hybrid Mixture of Experts (HyMoE) block inspired by~\cite{dai2024deepseekmoe} for the sequential recommendation task. Unlike conventional MoE that routes tokens to a sparse set of experts, we employ a dual-branch structure. It consists of a dense, shared expert branch and a sparse, specialized expert branch, ensuring stable learning while providing adaptive capacity.

Given a token representation $\mathbf{x} \in \mathbb{R}^{D}$, we first calculate a dense FFN output $\mathbf{y}_{\text{dense}}$. In parallel, $\mathbf{x}$ is passed through a lightweight router $\pi(\cdot)$, parameterized as a two-layer linear computation. Let $\mathcal{E}=\{f_1,\dots,f_E\}$ be the set of expert networks, the router produces $\pi(\mathbf{x}) \in \mathbb{R}^{E}$, where each dimension is a routing logit reflecting the affinity of the current token-context pair to one of the $E$ FFN experts.
Sparse gating weights are then obtained by retaining only the top-$k$ expert logits and applying a softmax over them: $\mathbf{g} = \text{Softmax}\!\big(\text{TopK}(\pi(\mathbf{x}))\big)$, where TopK is the selection of the K experts with the highest probabilities, $\mathbf{g}=\{g_1, ..., g_K\}$ represents the set of weights for the K experts.
The aggregated mixture of expert output is therefore computed as:
\begin{equation}
\mathbf{y}_{\text{MoE}} = \textstyle\sum_{i=1}^{K} g_i f_i(\mathbf{x}),
\end{equation}
where only TopK experts are active. This design brings two benefits: (1) dense branch guarantees a stable representation for every token, while the sparse branch adaptively increases expressiveness only when necessary; and (2) the training of the model is stabilized, which avoids the expert collapse.

\noindent\textbf{Adaptive Expert Fusion (AEF).} To effectively balance dense and sparse experts while ensuring stable training dynamics, instead of a fixed combination strategy that may fail to adapt to the evolving expertise of different components~\cite{dai2024deepseekmoe}, we design an Adaptive Expert Fusion (AEF) mechanism that dynamically combines the outputs of both branches. Our AEF employs a fusion parameter $\alpha$ combined with a progressive warm-up strategy:
\begin{equation}
\mathbf{y} = \mathbf{y}_{\text{dense}} + \alpha \cdot \mathbf{y}_{\text{MoE}}, \quad 
\alpha = \sigma(\alpha_{\text{param}}) \cdot w(t).
\end{equation}
Here, $\sigma(\cdot)$ is the sigmoid function applied to the learnable parameter $\alpha_{\text{param}}$, and $w(t)$ is the warm-up factor:
\begin{equation}
w(t) = \min(1.0, {t}/{T_{\text{warmup}}})
\end{equation}
where $t$ is the current step and $T_{\text{warmup}}$ is the warm-up duration.
{AEF} ensures gradual integration of expert knowledge while maintaining training stability. To avoid a few experts domination, we apply a load-balance loss to encourage uniform expert usage. Thus, the full training objective is:
\begin{equation}
\mathcal{L} = \mathcal{L}_{\text{ce}} + \lambda_{\text{lb}} \textstyle\sum_{e} \bar{g}_e \log \bar{g}_e,
\end{equation}
where $\bar{g}_e$ denotes the average gating probability of the expert $e$ on all items in the sequence, $\lambda_{\text{lb}}$ is the balancing parameter and $\mathcal{L}_{ce} = -\log \frac{\exp(\mathbf{y}^T \mathbf{r}_{+})}{\sum_{j} \exp(\mathbf{y}^T \mathbf{r}_j)}$ with $\mathbf{r}_{+}$ the embedding of the target element and $r_j$ as all embeddings. This regularization stabilizes training and promotes generalization by ensuring all experts contribute meaningfully to the predictions.

\begin{table}[t]
\resizebox{0.46\textwidth}{!}{
\begin{tabular}{l|l|cccc}
\toprule
Dataset & Method & HR@5 & HR@10 & NDCG@5 & NDCG@10 \\
\midrule
\multirow{6}{*}{ML-1M} & NARM & 0.1727 & 0.2565 & 0.1135 & 0.1404 \\
 & GRU4Rec & 0.1998 & 0.2815 & 0.1364 & 0.1628 \\
 & BERT4Rec & 0.1879 & 0.2868 & 0.1249 & 0.1568 \\
 & Mamba4Rec & 0.2106 & 0.3065 & 0.1447 & 0.1756 \\
 & \textbf{HyMoERec} & \textbf{0.2157} & \textbf{0.3103} & \textbf{0.1486} & \textbf{0.1790} \\
\midrule
\multirow{6}{*}{Beauty} & NARM & 0.0418 & 0.0630 & 0.0285 & 0.0352 \\
 & GRU4Rec & 0.0402 & 0.0635 & 0.0274 & 0.0349 \\
 & BERT4Rec & 0.0243 & 0.0402 & 0.0151 & 0.0202 \\
 & Mamba4Rec & 0.0511 & 0.0731 & 0.0362 & 0.0432 \\
 & \textbf{HyMoERec} & \textbf{0.0518} & \textbf{0.0756} & \textbf{0.0365} & \textbf{0.0442} \\
\bottomrule
\end{tabular}}
\caption{Recommendation performance on ML-1M and Beauty datasets. Bold scores represent the highest results.}
\label{tab:comprehensive_results_merged}
\vspace{-10pt}
\end{table}

\section{Experiments}

\textbf{Datasets and Setup.} We experiment on the MovieLens-1M and Beauty datasets. We set the number of experts $E$=4 and select top K=2 experts. $T_{\text{warmup}}$ is set to 500 steps, and $\lambda_{\text{lb}}$=0.02. We compare with NARM \cite{li2017neuralnarm}, GRU4Rec~\cite{jannach2017recurrentgru4rec}, BERT4Rec~\cite{sun2019bert4rec}, and Mamba4Rec~\cite{liu2024mamba4rec}.

\noindent\textbf{Results.} The results are shown in Tab. \ref{tab:comprehensive_results_merged}. On the ML-1M dataset, HyMoERec demonstrates superior performance with HR@5 of 0.2157, HR@10 of 0.3103, NDCG@5 of 0.1486, and NDCG@10 of 0.1790. These results represent superiority over the strongest baseline, Mamba4Rec, of 2.4\% in HR@5, 1.2\% in HR@10, 2.7\% in NDCG@5, and 1.9\% in NDCG@10. Compared to NARM and GRU4Rec, HyMoERec achieves 24.8\% and 7.9\% HR@5 improvements, demonstrating our effectiveness.
In the Beauty dataset, HyMoERec also shows substantial improvements compared to BERT4Rec, improving from 0.0243 to 0.0518 in HR@5 and from 0.0151 to 0.0365 in NDCG@5. These consistent improvements demonstrate the effectiveness and generalizability of our approach across different domains.

\section{Conclusion}

We introduce HyMoERec, a sequential recommendation framework that directly addresses the limitations of PFFNs by modeling the heterogeneity of user behaviors and item complexities. We propose the Hybrid Mixture of Experts (HyMoE) and Adaptive Expert Fusion (AEF) mechanisms, which provide a dynamic and stable approach to sequential recommendation. Experiments on MovieLens-1M and Beauty datasets show that HyMoERec consistently outperforms state-of-the-art baselines.

{
\vspace{1mm}
\scriptsize
\noindent{\textbf{Acknowledgements}}.
This research is supported by the Ministry of Education, Singapore, under its Academic Research Fund Tier 2 (Award No. MOE-T2EP20123-0015).
}

\bibliography{main}


\end{document}